\documentclass[12pt]{revtex4}
\usepackage{amssymb}
\usepackage{latexsym}
\usepackage{epsfig}
\begin{document}                             

\title{ Role of higher -dimensional evolving wormholes in the formation of a big rip singularity
 }

\author{M. R. Setare $^{1}$\footnote{rezakord@ipm.ir}, A.Sepehri $^{2}$\footnote{alireza.sepehri@uk.ac.ir}, }
\address{$^1$ Department of Science, Campus of Bijar,
University of Kurdistan, Bijar, Iran.\\ $^2$  Faculty of Physics,
Shahid Bahonar University, P.O. Box 76175, Kerman, Iran. }

\begin{abstract}
We study the evolutions of 4D universe on the M2-M5 BIon in the
thermal background. The BIon consists of a
D-brane and a parallel anti-D-brane which a wormhole connects them. When
the  branes and antibranes are far from each other and the brane's
spike and the antibrane's spike are well separated, the wormhole cannot be built.
However, when the branes and anti-branes are close to each other, a wormhole can be formed between them. Under this condition, there are many
ways for flowing  energy from transverse dimensions into our
universe. This energy overcomes all other forms of energy, such as the
gravitational repulsion  and  causes that our epoch is terminated at a Big-Rip singularity. We show that at this
point, universe would be disappeared and one black M2-brane is
formed. Finally, we examine our model against WMAP, Planck and BICEP2
experiments and get the ripping time. Comparing the model with experimental data, we obtain
$N\simeq 50$ and $n_{s}\simeq 0.96$, where \emph{N} and
$n_{s}$ are the number e-folds and the spectral index
respectively. These results could be located in $0.01 <
R_{Tensor-scalar } < 0.3$, where $R_{Tensor-scalar }$ is the
tensor-scalar ratio. In this epoch, the  Big Rip
singularity happens in a finite time $t_{rip}=31(Gyr)$ for WMAP and Planck experimnts
and $t_{rip}=28(Gyr)$ for BICEP2 experiment. By comparing this time with
the time of the Big Rip in brane-antibrane, we observe that the wormhole
in BIonic system causes that the destruction of the universe becomes faster.


\end{abstract}

 \maketitle
\section{Introduction}
Newly, some authors investigate about the probability that inflation
might provide a  method for the expansion of
Lorentzian wormholes of the Morris-Thorne type \cite{q1} to
macroscopic size. They argued that the throat and the proper length
of this wormhole expand \cite{q2}. Until now, the evolving
wormhole has been considered in many researchs
\cite{q3,q4,q5,q6,q7,q8}. For example, some  investigators studied
the effect of extra, decaying(in time) compact dimensions, existing
in the evolving Lorentzian wormhole metric  and observed  some
interesting results  for the cases of exponential and
Kaluza–Klein expansion \cite{q3}. Also, some researchers suggested
the logical solutions of General Relativity that explained the
evolving wormholes with a non-constant redshift function. They
described that the matter that threads these types of wormholes may not be
needed exotic \cite{q4}. In another papers, the characteristics of
evolving (2+1) and (3+1)-dimensional wormhole spacetimes within
the background of nonlinear electrodynamics are considered.  For the (3
+ 1)-dimensional spacetime, it is observed that the Einstein field
equation gives a contracting wormhole solution and the following
of the weak energy condition. Nevertheless, in the background of an
electric field, the latter provides a singularity at the throat,
however, for a pure magnetic field the solution is normal. For
the (2 + 1)-dimensional space-time, it is also observed that the physical
fields are singular at the throat \cite{q5,q6}. In another work,
authors investigated the probability of maintaining an evolving wormhole
via exotic matter constructed  of phantom energy. They discussed that this
exotic source can be protecting the present of evolving wormhole
spacetimes \cite{q7}. In one scenario, it was argued that for
positive cosmological constant, there exist some wormholes which enlarge
and, for negative cosmological constant, there exist some wormholes which
extend to a maximum value and then recollapse. Without the existence of
cosmological constant, the wormhole grows with constant
velocity, i.e without acceleration or deceleration. In 2+1
dimensions the enlarging wormholes always have an isotropic and
homogeneous pressure, relating to the time coordinate
\cite{q8}.  Another paper discussed about inhomogeneous and
anisotropic universe that terminates in a future singularity at a finite
value of the proper time. This model argued the evolving wormholes
for which the rate of enlargement is obtained by the phantom
energy, while the inhomogeneous and anisotropic component threads
and maintains the wormhole \cite{q9}.

Newly, some authors considered the probability of enlarging
wormholes in higher-dimensions, which is an important part
of modern theories of fundamental physics.  Since the Ricci scalar
is only corresponded to time in standard cosmological models, they
applied this property as a simplifying supposition. Also,
they argued about a particular type of wormhole solutions
relating to the choice of a spatially homogeneous Ricci
scalar. The probability of deriving solutions with ordinary and
exotic matter was investigated and they obtained a variety of solutions
including those in four dimensions that obey the null energy
condition (NEC) in specific time intervals  \cite{q10}.  Now, the
question arises that what is the origin of higher dimensional
wormhole in 4D universe? We will response to this question in M2-M5
BIonic configuration. It might be sensible to disregard the wormhole in the
ultraviolet where the  branes and antibranes are far from each other
and the brane's spike and the antibrane's spike are well distant, it is
possible that one BIon is built and expands  where the spikes of brane
and antibrane join to each other \cite{q11,q12}. In this system,
expansion of universe is affected by the wormhole between brane
and antibrane  and changes from non phantom state to phantom phase.
The wormhole energy density becomes large without bound and tends to
infinite in finite time, overcomes all other forms of energy, such
that the gravitational repulsion  and finally  causes that this era of universe terminates. When the universe becomes close to this
future singularity; all objects inside it like the Solar system, Earth etc will be torn up by the infinite phantom energy in a
Big Rip \cite{q13}. The finite time that this singularity in
BIonic configuratio happens is much sooner than the one in brane-antibrane
configuration in \cite{q14}.

The outline of the paper is as the following.  In section
\ref{o1}, we built four dimensional model of universe in M2-M5
BIonic system and argue that all cosmological parameters are functions of 
the wormhole's parameters between two branes. Also, in this
section, we study Big Rip singularity and obtain the
relation between ripping time and the size of wormhole in extra
dimension. In section \ref{o2}, we examine our model against the
observational data from Planck and WMAP experiments  and get
the ripping time. In section \ref{o3}, we study the signature
of finite temperature BIon in BICEP2 experiment. The last section is
devoted to summary and conclusion.

\section{ The destruction of four dimensional universe in M2-M5 BIon}\label{o1}
 In this section we will study the effects of the wormhole on the
evolution of scale factors, Hubble parameter and
other important parameters in FRW model of universe. We will
argue that these parameters are functions of the size of wormhole and are
affected by growing of wormhole in transverse dimensions. In our
mechanism, equation of state parameter in four dimensioal universe may evolve due
to flowing energy from transverse dimensions and change from higher
values of -1 (non-phantom state) to lower values (phantom phase).

To consider the BIon, we locate the M5-brane
world volume in eleven dimesional Minkowski space-time with below line-elements
\cite{q15,q16}:
\begin{eqnarray}
&& ds^{2} = -dt^{2} +  (dx^{1})^{2} + dr^{2} +
r^{2}d\Omega_{3}^{2} + \sum_{i=6}^{10}dx_{i}^{2}. \label{m1}
\end{eqnarray}
without background fluxes. Applying the standard angular coordinates
$(\psi,\phi,\omega)$ to declare the round three-sphere line-elements
\begin{eqnarray}
&& d\Omega_{3}^{2} = -d\psi^{2} +  sin^{2}\psi ( d\phi^{2} +
sin^{2}\phi d\omega^{2}). \label{m2}
\end{eqnarray}
We select the static gauge:
\begin{eqnarray}
&&t(\sigma^{a}) = \sigma^{1},\,x^{1}(\sigma^{1}) =
\sigma^{1},\,r(\sigma^{a})=\sigma^{2}\equiv \sigma  \nonumber \\&&
\psi(\sigma^{a}) = \sigma^{3} ,\, \phi(\sigma^{a}) = \sigma^{4},\,
\omega(\sigma^{a}) = \sigma^{5},\, x^{6}(\sigma^{a}) =
z(\sigma)\label{m3}
\end{eqnarray}
 Using this ansatz, the
induced line-elements on the effective fivebrane worldvolume is given by
\begin{eqnarray}
\gamma_{ab}d\sigma^{a}d\sigma^{b} = -(d\sigma^{0})^{2} +
(d\sigma^{1})^{2} + (1 + z'(\sigma)^{2})d\sigma^{2} +
\sigma^{2}(-d\psi^{2} +  sin^{2}\psi ( d\phi^{2} + sin^{2}\phi
d\omega^{2})) \label{m4}
\end{eqnarray}

 We constrain the two boundary conditions that $z(\sigma)\rightarrow 0$ for $\sigma\rightarrow \infty$ and $z'(\sigma)\rightarrow -\infty$ for $\sigma\rightarrow \sigma_{0}$, where $\sigma_{0}$ is the minimal two-sphere radius of the
system. After some mathematical calculations, we get \cite{q15,q16}:
\begin{eqnarray}
z_{\pm}(\sigma)= \int_{\sigma}^{\infty} ds
(\frac{F_{\pm}(s)^{2}}{F_{\pm}(\sigma_{0})^{2}}-1)^{-\frac{1}{2}}
\label{m5}
\end{eqnarray}
For finite temperature BIon, $F(\sigma)$ is obtained as:
\begin{eqnarray}
F_{\pm}(\sigma) = \sigma^{3}(\frac{1 + \frac{k^{2}}{\sigma^{6}}}{1
\pm \sqrt{1 - \frac{4q_{5}^{2}}{\beta^{6}}(1 +
\frac{k^{2}}{\sigma^{6}})}})^{3/2}(-2 +
\frac{3\beta^{6}}{2q_{5}^{2}}\frac{1 \pm \sqrt{1 -
\frac{4q_{5}^{2}}{\beta^{6}}(1 + \frac{k^{2}}{\sigma^{6}})}}{1 +
\frac{k^{2}}{\sigma^{6}}}) \label{m6}
\end{eqnarray}
where
\begin{eqnarray}
\beta = \frac{3}{4\pi T} ,\, q_{2} =
\sigma^{3}\frac{r_{0}^{3}}{2}sin\theta sinh2\alpha  ,\, q_{5} =
\frac{r_{0}^{3}}{2}cos\theta sinh2\alpha\label{m7}
\end{eqnarray}
with the definitions:
\begin{eqnarray}
&&cosh\alpha_{\pm} = \frac{\beta^{3}}{\sqrt{2}q_{5}}\frac{\sqrt{1
\pm \sqrt{1 - \frac{4q_{5}^{2}}{\beta^{6}}(1 +
\frac{k^{2}}{\sigma^{6}})}}}{\sqrt{1 +
\frac{k^{2}}{\sigma^{6}}}}\nonumber \\&& r_{0,\pm} =
\frac{\sqrt{2}q_{5}}{\beta^{2}}\frac{\sqrt{1 +
\frac{k^{2}}{\sigma^{6}}}}{\sqrt{1 \pm \sqrt{1 -
\frac{4q_{5}^{2}}{\beta^{6}}(1 + \frac{k^{2}}{\sigma^{6}})}}}
\nonumber \\&& tan\theta = \frac{k}{\sigma^{3}},\,q_{2} = k q_{5}
= -4\pi \frac{N_{2}}{N_{5}}l_{p}^{3}
 \label{m8}
\end{eqnarray}
In above results, $N_{2}$ and $N_{5}$ denote the number of M2 and
M5-branes and $q_{2}$ and $q_{5}$ refer to the charges of M2 and
M5-branes respectively and T marks  the temperature of BIon.  Joining a mirror solution to Eq.
(\ref{m5}), we build a wormhole configuration. The separation
distance $\Delta = 2z(\sigma_{0})$ between the N M5-branes and N
anti M5-branes for a choosen brane-antibrane wormhole configuration
is given by:
\begin{eqnarray}
\Delta = 2z(\sigma_{0})= 2\int_{\sigma_{0}}^{\infty}
ds(\frac{F(s)^{2}}{F(\sigma_{0})^{2}}-1)^{-\frac{1}{2}} \label{m9}
\end{eqnarray}

  At his stage, we want to build four dimensional world in thermal M2-M5 BIonic configuration. For this, it is necessary to consider the effect of the BIonic system on the four-
dimensional universe energy momentum tensor. We  focus on calculating
the EM tensor for $N_{5}$ M5-branes with an electric field on,
relating to k units of electric flux, at non-zero
temperature. We can derive this from a black M2-M5 bound state
geometry supposing that we are in the regime of large $N_{5}$ and
$g_{s}N$. We obtain, \cite{q11,q15,q16,q17},
 \begin{eqnarray}
&& T^{00}=\frac{\Omega_{3}\Omega_{4}}{16\pi
G}\beta^{3}\sigma^{3}\frac{F(\sigma)}{\sqrt{F^{2}(\sigma)-F^{2}(\sigma_{0})}}\frac{3cosh^{2}\alpha
+ 1}{cosh^{3}\alpha} \nonumber \\&& T^{ii}=
-\gamma^{ii}\frac{\Omega_{3}\Omega_{4}}{16\pi G}\beta^{3}\sigma^{3}
\frac{F(\sigma)}{\sqrt{F^{2}(\sigma)-F^{2}(\sigma_{0})}}\frac{1}{cosh^{2}\alpha},\,i=1,2,3
\nonumber \\
&&T^{zz}= \frac{\Omega_{3}\Omega_{4}}{16\pi G}\beta^{3}\sigma^{3}
\frac{F(\sigma)}{F(\sigma_{0})}\frac{3cosh^{2}\alpha
+ 1}{cosh^{3}\alpha}
 \label{m10}
\end{eqnarray}
This equation indicates that with growing temperature in BIonic system, the energy-momentum tensors grow. This is because that when spikes of  branes and antibranes are far from each other, wormhole isn't built and there isn't any way for flowing energy from transverse dimensions to our universe, however when two branes approach to each other and linked by a wormhole, temperature tends to large values.

The conservation law of energy-momentum shows that the tensors obtained in brane-antibrane system have the relation with ones corresponded to the four dimensional universe with the following equation:
 \begin{eqnarray}
&& T^{\mu\nu} = \frac{2}{\sqrt{-det g}}\frac{\delta S}{\delta g_{MN}}\frac{\delta g_{MN}}{\delta g_{\mu\nu}} = T_{MN}\frac{\delta g_{MN}}{\delta g_{\mu\nu}},
\label{m11}
\end{eqnarray}
where $T^{\mu\nu}$ denotes the energy-momentum tensor of four dimensional world in
eleven dimensional space-time with the metric of the form:
\begin{equation}
ds^{2} = ds^{2}_{Uni,1} + ds^{2}_{Uni,2} + ds^{2}_{wormhole},
\label{m12}
\end{equation}
Here
 \begin{eqnarray}
&& ds^{2}_{Uni1} = ds^{2}_{Uni2} = -dt^{2} + \tilde{a}(t)^{2}(dx^{2} + dy^{2} + dz^{2}),
\label{m13}
\end{eqnarray}
and for evolving metric, we apply the below metric \cite{q2,q10}:
\begin{eqnarray}
&& ds^{2}_{wormhole} = d\tau^{2} + R^{2}(\tau)\frac{dr^{2}}{1 -
c(r)} +  R^{2}(\tau)r^{2}d\phi^{2} \nonumber\\&& = d\tau^{2} +
D^{2}dr^{2} + R^{2}(\tau)r^{2}d\phi^{2} \label{m14}
\end{eqnarray}
   where R(t) denotes the scale factor, c(r) refers to an unknown dimensionless function, defined as c(r) = b(r)/r, where b(r) refers to the shape
   function and $D =\frac{R^{2}(\tau)}{1 - c(r)} $. In the remaining of this research, we suppose that $t_{1} = t_{2} = \tau =
   t$.  In this mechanism, we choose two four dimensional  universe that interact with each other via one three dimensional wormhole and build an eleven dimensional binary configuration .

To derive the energy- momentum tensor in this system, we apply the Einstein's field equation in
existence of fluid flow  that is given by:
\begin{equation}
{R_{ij}} - \frac{1}{2}{{\mathop{\rm g}\nolimits} _{ij}}R = k{T_{ij}}.
\label{m15}
\end{equation}
Puting the solution of this equation with the metric of Eq. (\ref{m12}) in the conservation law  of energy-momentum tensor in Eq. (\ref{m11}) and using Eq. (\ref{m10}) gives:
\begin{eqnarray}
&& kT_i^i = \frac{{5\ddot{ \tilde{a}}}}{\tilde{a}} + \frac{{{3{\dot{\tilde{a}}}^2}}}{{{\tilde{a}^2}}} + \frac{{5\dot{ \tilde{a}}\dot
D}}{{\tilde{a}D}} + \frac{{\ddot D}}{D}\nonumber\\
   && = -\gamma^{ii}\frac{\Omega_{3}\Omega_{4}}{16\pi G}\beta^{3}\sigma^{3}
 \frac{F(\sigma)}{\sqrt{F^{2}(\sigma)-F^{2}(\sigma_{0})}}\frac{1}{cosh^{2}\alpha},\,i=1,2,3
,  \nonumber\\
&&  kT_r^r =  kT_z^z = \frac{{6\ddot {\tilde{a}}}}{\tilde{a}} + \frac{{6{{\dot {\tilde{a}}}^2}}}{{{\tilde{a}^2}}}-
     \frac{(n-1)\ddot{R}}{R(t)} -
     \frac{(n-1)(n-2)\dot{R}^{2}}{2R^{2}}-\frac{(n-1)(n-2)c(r)}{2R(t)^{2}r^{2}}
\nonumber\\&&
   = \frac{\Omega_{3}\Omega_{4}}{16\pi G}\beta^{3}\sigma^{3}
   \frac{F(\sigma)}{F(\sigma_{0})}\frac{3cosh^{2}\alpha
   + 1}{cosh^{3}\alpha}
\nonumber\\
&& kT_{0}^{0} = \frac{{6{{\dot {\tilde{a}}}^2}}}{{{\tilde{a}^2}}} + \frac{(n-1)(n-2)c(r)}{2R^{2}(t)r^{2}} + \frac{(n-1)c'}{2R^{2}(t)r} + \frac{n(n-1)\dot{R}}{2R^{2}}
 \nonumber\\&& =  \frac{\Omega_{3}\Omega_{4}}{16\pi G}\beta^{3}\sigma^{3}\frac{F(\sigma)}{\sqrt{F^{2}(\sigma)-F^{2}(\sigma_{0})}}\frac{3cosh^{2}\alpha + 1}{cosh^{3}\alpha}
\label{m16}
\end{eqnarray}
in which  n is the number of dimensions which may be equal to eleven.
 The higher-dimensional stress-energy tensor will be supposed to be that of a perfect fluid and is given by:
\begin{equation}
T_i^j = {\mathop{\rm diag}\nolimits} \left[ { - p, - p, - p, - \bar{p}, - p, - p, - p, \rho } \right],
\label{m17}
\end{equation}
Here, $\bar{p}$ is the pressure in the transverse space-like dimension. In above relation,  the pressure in the transverse
dimension is different, in general, from the pressure in three dimensional space. Thus, this stress-energy tensor explains a
homogeneous, anisotropic perfect fluid in ten dimensions. Henc, the field equations are given by:
\begin{eqnarray}
&& p = \frac{{5\ddot{ \tilde{a}}}}{\tilde{a}} + \frac{{{3{\dot{\tilde{a}}}^2}}}{{{\tilde{a}^2}}} + \frac{{5\dot{ \tilde{a}}\dot
D}}{{\tilde{a}D}} + \frac{{\ddot D}}{D}\nonumber\\
   && = -\gamma^{ii}\frac{\Omega_{3}\Omega_{4}}{16\pi G}\beta^{3}\sigma^{3}
 \frac{F(\sigma)}{\sqrt{F^{2}(\sigma)-F^{2}(\sigma_{0})}}\frac{1}{cosh^{2}\alpha},\,i=1,2,3
,  \nonumber\\
&&  \overline{p} =  \frac{{6\ddot {\tilde{a}}}}{\tilde{a}} + \frac{{6{{\dot {\tilde{a}}}^2}}}{{{\tilde{a}^2}}}-
     \frac{(n-1)\ddot{R}}{R(t)} -
     \frac{(n-1)(n-2)\dot{R}^{2}}{2R^{2}}-\frac{(n-1)(n-2)c(r)}{2R(t)^{2}r^{2}}
\nonumber\\&&
   = \frac{\Omega_{3}\Omega_{4}}{16\pi G}\beta^{3}\sigma^{3}
   \frac{F(\sigma)}{F(\sigma_{0})}\frac{3cosh^{2}\alpha
   + 1}{cosh^{3}\alpha}
\nonumber\\
&& \rho = \frac{{6{{\dot {\tilde{a}}}^2}}}{{{\tilde{a}^2}}} +
\frac{(n-1)(n-2)c(r)}{2R^{2}(t)r^{2}} + \frac{(n-1)c'}{2R^{2}(t)r}
+ \frac{n(n-1)\dot{R}}{2R^{2}}
 \nonumber\\&& =  \frac{\Omega_{3}\Omega_{4}}{16\pi G}\beta^{3}\sigma^{3}\frac{F(\sigma)}{\sqrt{F^{2}(\sigma)-F^{2}(\sigma_{0})}}\frac{3cosh^{2}\alpha + 1}{cosh^{3}\alpha}
\label{m18}
\end{eqnarray}
where we put the higher-dimensional
coupling constant equal to one, k = 1. Equation (\ref{m18}) help us to describe all properties of the current world in terms of phenomenologcal events in BIonic configuration. These equations impose a relation between the pressure and density in our universe with wormhole  and show that any increase or decrease in these parameters is due to wormhole evolution in transverse dimension.

 Obtaining the solutions of  equations (\ref{m6}, \ref{m9} and \ref{m18}) simultaneously  and doing some mathematical calculations for old world, we derive a consistent solution
for the scale factors, Hubble parameter $H=\frac{\dot{\tilde{a}}}{\tilde{a}}$ and the shape
function of wormhole and BIonic coordinates :
\begin{eqnarray}
&& H = \frac{\dot{\tilde{a}}}{\tilde{a}}\sim \frac{A}{t_{0} - t}\nonumber \\&& A=(\frac{\Omega_{3}\Omega_{4}}{16\pi G}\beta^{3}\frac{(3cosh^{2}\alpha + 1)}{cosh^{3}\alpha})^{1/2}\nonumber \\&& \tilde{a} = (t_{0} - t)^{A} \label{m19}  \\\nonumber \\&&\sigma^{2} = \sigma_{0}^{2} + (t_{0} - t)^{4}\nonumber \\&& R \sim n(n-1)A^{-1}(t_{0} - t)^{1/2} \label{m20}
 \\\nonumber  \\
&& c=\frac{b(r)}{r}\sim \frac{n^{2}(n-1)}{n-2}r \nonumber\\
&& r\sim (t_{0} - t), \label{m21} \\&& \nonumber\\&& \Delta \sim
\frac{3cosh^{2}\alpha +1}{4cosh^{3}\alpha}(t-t)^{1/2}. \label{m22}
\end{eqnarray}

Equations (\ref{m19}, \ref{m20} and \ref{m21}) declare that both universe scale factor and wormhole scale factor are related to the finite temperature of BIon and charge of M5-branes. With growing temperature,  the energy of system grows and the effect of interaction between M5-branes may be seen in scale factors. Also, this equation explains that if A be smaller than zero, three dimensional scale factor and Hubble parameter at an special time (t = $t_{0}$=$t_{rip}$) are infinity. Previously, investigators have believed that if three-dimensional equation of state parameter is lower than negative one, Dark energy is phantom energy and at Big Rip singularity, components of which universe is built will be torn up; whereas our results show that hapening of this singularity is corresponded to some events in transverse dimension. On the other hand, as may be bserved from equations (\ref{m20} and \ref{m21}), the value of the scale factor in transverse dimension and the shape function of wormhole would be zero at $t = t_{0}$ and as a conclusion  Big Rip singularity won't occur in transverse dimension.

The equation (\ref{m22}) has some interesting findings which may be applied to describe the reasons for hapenin of Big Rip singularity in present epoch of universe. According to these results, when two M5-branes are placed at two pints with a large separation distance from
each other ($\Delta=\infty$ and t=0), the wormhole parameters are almost zero, while closing the two together, spikes of two M5-branes join, wormhole is built and  the value
of it's parameters grows. Under this condition, brane-antibrane system vanish and consequently one singularity occurs in our four dimensional universe. Another interesting result that  this equation gives us is that time of this singularity is a function of  the initial separation distance between two branes.

At this stage, the main question arises that what is the fate of
universe-wormhole configuration after the Big Rip singularity? To response
to this question, we study the matching of finite temperature
BIon and black M2-M5 at transition point\cite{q16}. The M2
brane has the same charge $Q_{2}$ and the same temperature T as
the M2-M5 configuration. A perturbative extension of the tension around
the extremal limit yields \cite{q16}:

\begin{eqnarray}
\frac{M_{M2-brane}}{L_{x^{1}}L_{z}}= Q_{2}(1+
\frac{\sqrt{q_{2}}}{3\sqrt{2}\beta^{3}}+\frac{5q_{2}}{2^{6}\beta^{6}})
\label{m23}
\end{eqnarray}

On the other hand, for finite temperature BIon, we obtain \cite{q16}:
\begin{eqnarray}
&& \frac{dM_{BIon}}{L_{x^{1}}dz} =
\frac{\Omega_{3}\Omega_{4}}{16\pi G}\beta^{3}\sigma^{3}
   \frac{F(\sigma)}{F(\sigma_{0})}\frac{3cosh^{2}\alpha
   + 1}{cosh^{3}\alpha}
   \label{m24}
  \end{eqnarray}
and consequently, for small temperature, we get:
\begin{eqnarray}
&& \frac{dM_{BIon}}{L_{x^{1}}dz}=
Q_{2}(\sqrt{1+\frac{\sigma^{6}}{k^{2}}} +
\frac{5q_{2}^{2}}{6\beta^{6}}\frac{(1+\frac{\sigma^{6}}{k^{2}})^{3/2}}{\sigma_{0}^{6}} + \frac{11q_{2}^{4}}{8\beta^{12}}\frac{(1+\frac{\sigma^{6}}{k^{2}})^{5/2}}{\sigma_{0}^{12}}
)\label{m25}
\end{eqnarray}

If we compare the mass densities for BIon system with the the mass density for  black M2-brane, we observe that in order for the thermal
BIon at $\sigma = \sigma_{0}$  is similar to the black M2-brane, $\sigma_{0}$
should have the below dependence on the temperature \cite{q16}:
\begin{eqnarray}
&& \sigma_{0} = \frac{q_{2}^{1/4}}{\beta^{1/2}}(1.234 -
0.068\frac{q_{2}^{1/2}}{\beta^{3}}) \label{m26}
\end{eqnarray}
Now, applying (\ref{m19}, \ref{m20}, \ref{m21} and \ref{m26}), we may
obtain the scale factor in four dimensional universe and the shape
function of wormhole in terms of temperature:
\begin{eqnarray}
&& \tilde{a} = (\sigma^{2} - \sigma_{0}^{2})^{A/4} \sim
(\sigma^{2} - \frac{q_{2}^{1/2}}{\beta^{1/2}}(1.234 -
0.068\frac{q_{2}^{1/2}}{\beta^{3}})^{2})^{A/4}
 \nonumber\\&&
  R= \frac{n^{2}(n-1)}{n-2}(\sigma^{2} - \frac{q_{2}^{1/2}}{\beta^{1/2}}(1.234 -0.068\frac{q_{2}^{1/2}}{\beta^{3}})^{2})^{1/4}
 \nonumber\\&&
 b(r)\sim(\sigma^{2} - \frac{q_{2}^{1/2}}{\beta^{1/2}}(1.234 - 0.068\frac{q_{2}^{1/2}}{\beta^{3}})^{2})^{1/2}
\label{m27}
\end{eqnarray}
 These results indicate that as the wormhole expands, there exist more ways for flowing energy from transverse dimension to other four dimensions. Growing this energy leads to the destruction of universe and the building of the black M2-brane.

\section{Considering the signature of finite temperature M2-M5 BIon in Planck and WMAP9 data}\label{o2}
In previous section, we suggest a model that let us to study the FRW universe in M2-M5 BIonic configuration. Here, we test the correctness of our mechanism with experimental data and derive some important parameters like ripping time. Previously, eight possible asymptotic solutions for cosmological dynamics have been suggested
\cite{q18}. Three of these solutions give the non-inflationary scale factor and another three ones
of solutions yield de Sitter, intermediate and power-low inflationary expansions. Also, two
ones of these solutions yield asymptotic expansion with scale factor ($a = a_{0}exp(A[lnt]^{\lambda})$.
This type of inflation is called logamediate inflation \cite{q19}. Our model allows us to consider
BIonic inflationary mechanism in the scenarios of  logamediate
inflation with ${\lambda}=1$. Under this condition, the mechanism is changed to power-law inflation. Applying (\ref{m22}), the number of e-folds may be obtained as:
\begin{eqnarray}
&&N=\int_{t_{B}}^{t} H dt=
|A|[ln\frac{t_{rip}-t_{B}}{t_{rip}-t}]\nonumber \\&&
=(\frac{\Omega_{3}\Omega_{4}}{16\pi
G}\beta^{3}\frac{(3cosh^{2}\alpha +
1)}{cosh^{3}\alpha})^{1/2}[ln\frac{t_{rip}-t_{B}}{t_{rip}-t}]
\label{m28}
\end{eqnarray}
where $t_{B}$ refers to the beginning time of inflation e.

In Fig.1, we show the number of e-folds \emph{N} for BIonic
configuration as a function of the t where t is the age of universe. In
this figure, we select A=-91, , $T = 2^{0}k$, $t_{B}=0$ and
$t_{rip}=31(Gyr)$. We observe that \emph{N}=50 leads to
$t_{universe}= 13.5(Gyr)$. This result is in agreement with 
both Planck and WMAP9 experiments \cite{q20,q21,q22}. Obviously,
the number of e-folds \emph{N} is much bigger for older universe.
This is because, as the age of universe grows, the distance
between spikes of branes becomes smaller, wormhole is built and
more energy dissolves into the universe.

\begin{figure}\epsfysize=10cm
{ \epsfbox{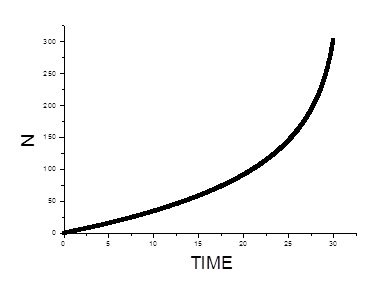}}\caption{The number of e-folds \emph{N} in
BIonic inflation scenario as a function of $t$ for  A=-91, $T =
2^{0}k$, $t_{B}=0$ and $t_{rip}=31(Gyr)$. } \label{1}
\end{figure}

Another parameters that help us to examine our mechanism with experimental observations are the power-spectrum of scalar and tensor perturbations which are given by \cite{q23}:
\begin{eqnarray}
&&\Delta_{R}^{2}=- (\frac{\Gamma^{3}T^{2}}{36(4\pi)^{3}})^{\frac{1}{2}}\frac{H^{\frac{3}{2}}}{\dot{H}} \nonumber\\
&&\Delta_{T}^{2}= \frac{2H^{2}}{\pi^{2}}, \label{m29}
\end{eqnarray}
 Applying these parameters, we can obtain the tensor-scalar ratio as \cite{q23}:
\begin{eqnarray}
&& R_{Tensor-scalar }= -\frac{\Delta_{T}^{2}}{\Delta_{R}^{2}} =
-(\frac{144(4\pi)^{3}}{\Gamma^{3}\pi^{4}
T^{2}})^{\frac{1}{2}}\dot{H}H^{\frac{1}{2}}, \label{m30}
\end{eqnarray}
In Fig.2, we present the tensor-scalar ratio $R_{Tensor-scalar }$ in
BIonic configuration as a function of the age of universe. In this
figure, we select A=-91, $t_{B}=0$, $T = 2^{0}k$, c=.01, $\Gamma =
.3$ and $t_{rip}=31(Gyr)$. We conclude that $R_{Tensor-scalar }$=0.021
leads to $t_{universe}= 13.5(Gyr)$. This result is in agreement with
both Planck and WMAP9 experiments \cite{q20,q21,q22}. It is clear that the
tensor-scalar ratio $R_{Tensor-scalar }$ is much bigger for older
universe. The reason for this is that when two old brane universes
become close together, the size of wormhole grows and it's effect
can be seen in observational experiments.

\begin{figure}\epsfysize=10cm
{ \epsfbox{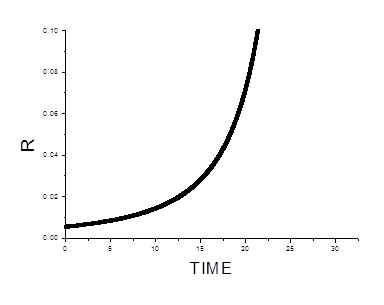}}\caption{The tensor-scalar ratio
$R_{Tensor-scalar }$ in BIonic inflation scenario as a function of
$t$ for A=-91, $T = 2^{0}k$, $t_{B}=0$, C=.01, $\Gamma = .3$ and
$t_{rip}=31(Gyr)$. } \label{2}
\end{figure}

Finally, we test our model with the Scalar spectral index which is given by \cite{q23}:
\begin{equation}
n_{s}-1=-\frac{d ln\Delta_{R}^{2}}{d ln k}=\frac{3}{2}\varepsilon - (\frac{r'}{4r})\eta
\label{m31}
\end{equation}
Here  k is co-moving wavenumber, $r=\frac{\Gamma}{3H}$ and $\varepsilon$ and $\eta$ are slow-roll parameters of the BIonic  inflation which are obtained as:
\begin{eqnarray}
&& \varepsilon =- \frac{1}{H}\frac{d ln H}{dt} \nonumber\\
&& \eta = -\frac{\ddot{H}}{H\dot{H}}, \label{m32}
\end{eqnarray}
In Fig.3, we present the Scalar spectral index $\emph{n}_{s}$ for
BIonic configuration as a function of the  age of universe. In this
figure, we select A=-91 and $t_{B}=0$, $\Gamma = .3$ and $T =
2^{0}k$. Comparing this figure with figures(2,3), we observe that
$N\simeq 50$ case leads to $n_{s}\simeq 0.96$. This standard case
is seen in $0.01 < R_{Tensor-scalar } < 0.22$, which is
in agreement with both Planck and WMAP9 experiments \cite{q20,q21,q22}. In this epoch, the finite time that Big Rip singularity happens is
$t_{rip}=31(Gyr)$.
\begin{figure}\epsfysize=10cm
{ \epsfbox{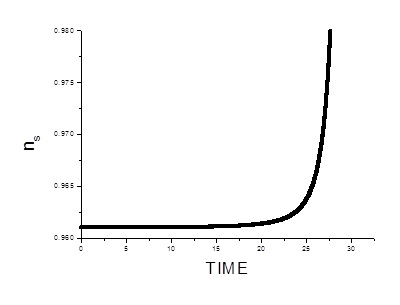}}\caption{The Scalar spectral index
$\emph{n}_{s}$ in BIonic inflation configuration as a function of $t$
for A=-91, $T = 2^{0}k$, $t_{B}=0$, $\Gamma = .3$ and
$t_{rip}=31(Gyr)$. } \label{3}
\end{figure}
\section{Considering the signature of finite temperature M2-M5 BIon in BICEP2 results}\label{o3}
The BICEP2  experiments \cite{q24} has reported newly the detection of primordial B-mode polarization in the
cosmic microwave background (CMB), asserting an indirect
observation of gravitational waves. Such results from
 BICEP2 suggest a rare opportunity to directly examine
theoretical models, containing inflation. The tensor fluctuations
in the cosmic microwave background (CMB) temperatures at large
angular scales are bigger than those suggested for inflationary
models based on Einstein gravity. Also, the ratio of
tensor-to-scalar perturbations announced by BICEP2 collaboration,
$R_{Tensor-scalar } = 0.2^{+0.07}_{-0.05}$ (or $R_{Tensor-scalar }
= 0.16^{+0.06}_{-0.05}$ after subtracting an estimated
foreground), is bigger than the bounds $R_{Tensor-scalar } < 0.13$
and $R_{Tensor-scalar } < 0.11$ announced by Planck\cite{q20} and
WMAP9 experiments \cite{q21,q22}. Fortunately, our mechanism is in agreement with all results of three
experiments.
 This is because that tensor fluctuations is related to the finite temperature and number of branes in BIonic configuration and can contain both smaller quantities of 0.1 and larger quantities of 0.19.
 From (\ref{m8} and \ref{m19}), it is observed that for small
temperatures, the function A has the below form:
\begin{eqnarray}
&& A=(\frac{\Omega_{3}\Omega_{4}}{16\pi G}\beta^{3}\frac{(3cosh^{2}\alpha + 1)}{cosh^{3}\alpha})^{1/2}
\nonumber\\
&& \simeq q_{5}(1 + (\frac{N_{2}}{N_{5}})^{2})(2 +
\frac{4q_{5}^{2}}{\beta^{6}}(1+(\frac{N_{2}}{N_{5}})^{2}))
\label{m33}
\end{eqnarray}
where  $N_{5}$ and $N_{2}$ denote the number of M5-branes and
M2-branes and T refers to the temperature.
 Applying this equation, we can obtain the tensor-scalar ratio in terms of NT :
\begin{eqnarray}
&& R_{Tensor-scalar } \sim
 q_{5}^{3/2}(1 + (\frac{N_{2}}{N_{5}})^{2})^{3/2}(2 +
\frac{4q_{5}^{2}}{\beta^{6}}(1+(\frac{N_{2}}{N_{5}})^{2}))^{3/2}
(\frac{1}{t-t_{rip}})^{5/2} \label{m34}
\end{eqnarray}
  Equation (\ref{m34}) indicates that any increase or decrease in the number of M-branes and temperature can lead to a significant change in the tensor-scalar ratio.
  For example, if temperature or number of branes is $10^{\frac{2}{3}}$ higher than the value estimated in previous section,
   the ratio of tensor-to-scalar perturbations would be $R_{Tensor-scalar } \sim 0.20$ which is in agreement with the value measured by BICEP2 experiments.

 In Fig.4, we present the tensor-scalar ratio $R_{Tensor-scalar }$ for
 BIonic
configuration as a function of the age of universe.
 In this figure, we choose A=-35, $t_{B}=0$, $T = 2^{0}k$, $\frac{N_{2}}{N_{5}} \sim 10^{8/3}$, c=.01 and $t_{rip}=28(Gyr)$. We observe that $R_{Tensor-scalar }=0.20$ leads to $t_{universe}= 13.5(Gyr)$. This ratio is in agreement with  BICEP2  experiment \cite{q24}. Thus our result are in agreement with all experimental data from WMAP, Planck and BICEP2 observations and thus our model works.

\begin{figure}\epsfysize=10cm
{ \epsfbox{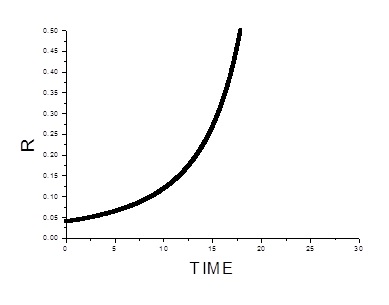}}\caption{The tensor-scalar ratio
$R_{Tensor-scalar }$ in BIonic inflation scenario as a function of
$t$ for A=-55, $T = 2^{0}k$, $\frac{N_{2}}{N_{5}} \sim 10^{4/3}$,
$t_{B}=0$, C=.01 and $t_{rip}=28(Gyr)$. } \label{2}
\end{figure}

\section{Summary and Discussion} \label{sum}
In this paper, we built four dimensional universe in thermal M2-M5 BIon and indicate that the energy density, slow-roll, Number of e-fold and
perturbation parameters can be obtained  in terms of the shape function and the location of the  throat of the wormhole that makes a bridge between brane and antibrane
in BIonic system. When the  branes and antibranes are far from each other and the brane's spike and the antibrane's spike are well seperated, the wormhole may be disregarded,
 however as  the spikes of brane and antibrane
join each other, one wormhole would be built.
 According to our findings, when the wormhole parameters grow, the Number of e-fold, the spectral index and the tensor-scalar ratio  grow and tend to
 infinity at Big Rip singularity. This is because that when the wormhole expands, the effect of interaction between branes and anti-branes on the four dimensional universe expansion becomes more effective, because for bigger wormholes, there are more ways for flowing energy from transverse dimensions to other four dimensions.
 Growing this energy causes that the  universe disappears and  the black M2-brane is built. We observed that $N\simeq 50$ case causes to $n_{s}\simeq 0.96$.
 These results may be located in $0.01 < R_{Tensor-scalar } < 0.3$, which is in agreement with experimental data \cite{q20,q21,q22,q24}.
  In this epoch, the finite time that Big Rip singularity happens is $t_{rip}=31(Gyr)$ for WMAP and Planck experiments and $t_{rip}=28(Gyr)$ for BICEP2 experiment.
   This time is much sooner of ripping time in brane-antibrane configuration in Ref \cite{q14}.

 \end{document}